\documentstyle[pra,aps,epsf]{revtex}
\begin{document}
\draft
\title{Antiproton-Hydrogen annihilation at sub-kelvin temperatures}
\author{A. Yu. Voronin}
\address{P. N. Lebedev Physical Institute\\
53 Leninsky pr.,117924 Moscow, Russia}
\author{J. Carbonell}
\address{Institut des Sciences Nucl\'eaires\\
53, Av. des Martyrs, 38026 Grenoble, France}
\maketitle

\begin{abstract}

The main properties of the interaction of ultra low-energy antiprotons ($%
E\le10^{-6}$ a.u.) with atomic hydrogen are established. They include the
elastic and inelastic cross sections and Protonium (Pn) formation spectrum. The
inverse Auger process ($Pn+e \rightarrow H+\bar{p}$) is taken into account
in the framework of an unitary coupled-channels model. The annihilation
cross-section is found to be several times smaller than the predictions 
made by the
black sphere absorption models. A family of $\bar{p}H$ nearthreshold
 metastable states is predicited. The dependence of Protonium
formation probability on the position of such nearthreshold S-matrix
singularities is analysed. An estimation for the $H\bar{H}$ annihilation
cross section is obtained.
\end{abstract}

\pacs{:  03.65.Nk; 34.10.+x, 11.80.J, 36.10.Dr}


\section{Introduction}

The unique features of the LEAR (low-energy antiproton ring) facility at
CERN made recently possible the synthesis of few antihydrogen atoms \cite
{ANTIH}. This effort would be pursued with the antiproton deccelerator (AD)
project \cite{AD} and would make possible the storage of sensible amounts of
antihydrogen. This project reinforces the already active interest to
investigate several theoretical and experimental problems in the physics of
antimatter \cite{Gabr1,Gabr2,Munich 92,Charlton 94,LEAP94}.

In view of storing antimatter in traps, it would be interesting to have some
theoretical calculations of the rate at which antiprotons ($\bar{p}$) and
antihydrogen ($\bar{H}$) annihilate with the residual gas. This process
should be evaluated at subkelvin temperatures, the optimal energy domain for
an effective synthesis and trapping of antimatter.

 From a theoretical standpoint, a specific feature of systems like $H +\bar{p}
$ or $H +\bar{H}$, containing pairs of unlike charged heavy particles ($p$
and $\bar{p}$), is the possibility of rearrangement followed by Protonium
(Pn) formation. Indeed even at zero \={p} kinetic energy Protonium can be
produced in states with principal quantum number $n \leq 30$.

Since the pioneer work of E. Fermi and E. Teller \cite{Fermi} in 1947, this
problem has been treated by several authors \cite{Morgan,Junk,SWS} in the
energy range going from a fraction to tens a.u.. The usual approach is based
on the following two assumptions: i) the separation of electronic and
nuclear motion and ii) the classical treatment of the antiproton dynamics.
The aim of the present work is to provide a correct description of the
ultra-low energy limit, i.e. $T_{\bar{p}}<10^{-6}$ a.u., an energy domain in
which the above mentioned assumptions are no longer valid. As a consequence
we are definitely faced to a quantum three- or four-body problem.

We will consider in what follows the annihilation of ultra slow antiprotons
with atomic hydrogen in the framework of an unitary coupled channel
approach. This process will be identified to the free Pn formation:
\begin{equation}  \label{react}
\bar{p}+H \rightarrow Pn^{*} + e
\end{equation}
as far as the direct annihilation at such low energies can be neglected and
that the further evolution of Pn states will result into
annihilation.

In our treatment the Protonium formation as well as the virtual
rearrangement process, i.e. the inverse reaction of (\ref{react}):
\begin{equation}  \label{invert}
\bar{p} + H \rightarrow Pn^{*} + e \rightarrow \bar{p} + H
\end{equation}
will be properly taken into account.

We will show that the energy dependence of the inelastic reaction
 probability is determined by a
rich spectrum of nearthreshold S-matrix singularities, corresponding to $H%
\bar p$ nearthreshold metastable states generated by the long range
charge-dipole interaction. Finally we will give an estimation of the $H\bar H
$ annihilation cross-section in the energy range from $10^{-8}$ to $10^{-4}$
a.u.


\section{The formalism}

The adequate formalism for the three-body problem are the Faddeev equations
\cite{Fad}, according to which three possible asymptotic clusters have to be
explicitly described. In the case of slow antiprotons scattering on hydrogen
only two of them are physically important. The first one is the $(pe)\bar{p}$
cluster, which corresponds to the elastic channel $H+\bar{p}\rightarrow H+%
\bar{p}$. The second one is the $(p\bar{p})e$ cluster, corresponding to the
Protonium formation channels $H+\bar{p}\rightarrow Pn^{*}+e$.

A direct solution of the Faddeev equations for such a problem is made
difficult by the big number of open channels containing fast oscillating
asymptotics. In this section we will develop a formalism which enables to
take into account the correct behavior of the three-body wavefunction in
both mentioned asymptotic clusters and benefit from the small value of
the electron-antiproton mass ratio. We will show that this approach gives
the scattering observables with an accuracy of the order of
$\sim 10\%$ by using a very limited number of channels.
In the same time it provides a transparent physical understanding
of the low energy three-body dynamics in the $\bar{p}H$ reaction,
which was the main aim of our study.


\subsection{The coordinate system}

The Jacobi coordinates for the 3-body problem are connected with the
possible different asymptotic clusters. The coordinates corresponding to the
elastic channel are defined by
\begin{eqnarray}  \label{Jfr}
{\bf r}&=&{\bf r}_{e}-{\bf R}_{p}  \nonumber \\
\\
{\bf R}&=& {\bf R}_{\bar{p}}-{\frac{m_{e}{\bf r}_e+M_{p}{\bf R}_p}{m_{e}+M_{p}}}  
\nonumber
\end{eqnarray}
where ${\bf {r}_{e}}$, ${\bf {R}_{p}}$ and ${\bf {R}_{\bar{p}}}$ are
respectively the electron, proton and antiproton coordinates, and $m_{e}$ $%
M_{p}$ are the electron and proton mass.

It turns out that this frame is also convenient for describing the Protonium
production channels. Indeed, the $p$-$\bar p$ distance ${\bf {\rho }={R_p-R_{%
\bar p}}}$, coincides within ${\bf {r}}(m_e/M_p{\bf })$ with ${\bf {R}}$
in equation (\ref{Jfr}). We will show that knowing the
three-body wavefunction at ${\bf {r}\ll {\rho }(}M_p/m_e{\bf )}$ is enough to
get a good approximation of the scattering observables. Thus we can
substitute ${\bf {\rho }}$ by ${\bf {R}}$ and take into account, if
necessary, the difference between ${\bf {\rho }}$ and ${\bf {R}}$ in a
perturbative way. We will use hereafter the unique coordinate system (\ref
{Jfr}).


\subsection{Three-body wavefunction}

The three-body wavefunction is represented as a sum of two components,
corresponding to the two considered clusters:
\begin{equation}  \label{Phicomp}
\Phi({\bf {r},{R}}) = \Phi_{1}({\bf {r},{R}}) +\Phi_{2}({\bf {r},{R}})
\end{equation}

Component $\Phi_{1}$ is supposed to describe the elastic channel and can be
written as
\begin{equation}  \label{Phi10}
\Phi_{1}({\bf {r},{R})}= \sum_{\alpha}\phi_{\alpha}({\bf r})\chi_{\alpha}(%
{\bf R})
\end{equation}
where $\alpha$ is the set of quantum numbers labeling the Hydrogen atomic
states $\phi_{\alpha}$ as well as the corresponding antiproton wavefunction $%
\chi_{\alpha}$. We will denote from now by channel each term in the
three-body wavefunction expansion like (\ref{Phi10}).

For incident antiproton energy much smaller than the first Hydrogen
excitation threshold, it is convenient to select from (\ref{Phi10}) only the
contributions which do not vanish in the asymptotics. This gives the simple
form:
\begin{equation}
\Phi _1({\bf {r},{R})}=\phi _{1s}({\bf {r}})\chi ({\bf {R}})  \label{Phi1}
\end{equation}
It is useful to introduce a projection operator $\hat P$, which acts in the
three-body states space and projects on the subspace of Hydrogen states
corresponding to open channels:
\begin{equation}
\hat P=\sum_{nlm}|\phi _{nlm}\rangle \langle \phi _{nlm}|\otimes \hat 1
\label{hatP}
\end{equation}
In our case the sum is limited to the 1s state only. Component $\Phi _1$ is
then written as:
\begin{equation}
\Phi _1=\hat P\Phi  \label{Phi1proj}
\end{equation}
The second component $\Phi _2$ describes all the remaining channels and can
be written in the form:
\begin{equation}
\Phi _2=(\hat 1-\hat P)\Phi  \label{Phi2proj}
\end{equation}
This means that all the electron states except 1s contribute into $\Phi _2$
ensuring the orthogonality of both components. The $\Phi _2$ component
contains terms which correspond to the Protonium formation channels. In
order to explicitly take into account the asymptotic behavior of the cluster
$(p\bar p)e$, $\Phi _2$ is expanded in a complete set of the $(p\bar p)$
eigenfunctions $f_\beta ({\bf {R})}$:
\begin{equation}
\Phi _2({\bf {r},{R}})=\sum_\beta g_\beta ({\bf {r}})f_\beta ({\bf {R}})
\label{phi2exp}
\end{equation}
where $g_\beta ({\bf {r})}$ are unknown expansion coefficients representing
the electron wavefunctions in the channels characterized by Protonium
quantum numbers $\beta $.

Let us remark here that at this level no approximation has been done.
In particular the truncation done in the choice (\ref{Phi1}) of
$\Phi_1$ is balanced in the functions $g_{\beta}$ of
the second component.

\subsection{Equations}

The Schrodinger equation for a three-body vector state $|\Phi >$ reads:
\begin{equation}
(\hat H_{ep}+\hat H_{p\bar p}^{ex}+\hat W_{e\bar p}^{ex}-E)|\Phi >=0
\label{opeq}
\end{equation}
with:
\begin{equation}\label{HHam}
\hat H_{ep}=-{\frac 1{2m_e}}{\bf \Delta }_r-{\frac 1r}
\end{equation}
the Hydrogen Hamiltonian,
\begin{equation}\label{PHam}
\hat H_{p\bar p}^{ex}=-{\frac 1{2M}}{\bf \Delta }_R-{\frac 1{|{\bf {R}+{r}}%
(m_e/M_p)|}}
\end{equation}
the Protonium Hamiltonian and
\begin{equation}\label{Vepb}
\hat W_{e\bar p}^{ex}={\frac 1{|{\bf {R}-{r}}(1-m_e/M_p)|}}
\end{equation}
the electron-antiproton interaction potential.
$M$ is the $\bar{p}H$ reduced mass which, neglecting $m_e/M_p$ terms,
will be hereafter approximated by the Protonium reduced mass
$M\approx M_p/2$,
$E=\varepsilon _B+E_{\bar p}$ the total energy,
$\varepsilon_B$
 the Hydrogen ground state energy and $E_{\bar p}$ the center of mass
energy of incident antiproton. Note that all the spin degrees of freedom are
neglected. We will first neglect the term ${\bf r(}m_e/M_p)$ and substitute
the exact $\hat H_{p\bar p}^{ex}$ and $\hat W_{e\bar p}^{ex}$ by the
approximations:
\begin{eqnarray*}
\hat H_{p\bar p} &=& -{\frac 1{2M}}{\bf \Delta }_R-{\frac 1R} \cr
\hat W_{e\bar p} &=& {\frac 1{|{\bf {R}-{r}|}}}
\end{eqnarray*}
Using (\ref{Phi1proj}) and (\ref{Phi2proj}) we obtain the following coupled
equations for the components $|\Phi _i>$:
\begin{eqnarray}
(\hat H_{p\bar p}+\hat P\hat W_{e\overline{p}}\hat P-E_{\bar p})|\Phi
_1\rangle +\hat P\hat W_{e\overline{p}}(1-\hat P)|\Phi _2\rangle &=&0
\nonumber  \\
&&  \label{eqsf} \\
(\hat H_{ep}+\hat H_{p\bar p}+(1-\hat P)\hat W_{e\overline{p}}(1-\hat P%
)-E)|\Phi _2\rangle +(1-\hat P)\hat W_{e\overline{p}}\hat P|\Phi _1\rangle
&=&0  \nonumber
\end{eqnarray}
We used here the fact that $\hat P$ commutes with $\hat H_{ep}$
and $\hat H_{p\bar p}$.

The corresponding equations for the antiproton $\chi (R)$ and electron
wave-functions $g_\beta (r)$ could be obtained by substituting (\ref{Phi1})
and (\ref{phi2exp}) into (\ref{eqsf}). Due to the choice of the wavefunction
components the solution of such a coupled equations system will correctly
describe the asymptotic behavior of the three-body system. This procedure
will however remain formal, for it leads to an infinite set of coupled
channels, including the closed ones, characterized by a continuous $p\bar p$
momentum variable. To construct an equation system suitable for practical
calculations, we should first analyze the contribution of different channels
in the expansion (\ref{phi2exp}). In particular, the one coming from the
continuous spectrum. For such a purpose we represent $\Phi _2$ as a sum of
two components:
\begin{eqnarray}
\Phi _2=\Phi _2^d+\Phi _2^p  \label{D&P}
\end{eqnarray}
where
\begin{equation}
\Phi _2^d({\bf {r},{R})=}\hat F\Phi _2{\bf ({r},{R})=}\sum_\beta
^{n_{max}}g_\beta {\bf ({r})}f_\beta {\bf ({R})}  \label{D}
\end{equation}
\begin{equation}
\Phi _2^p({\bf {r},{R})=(}1-\hat F{\bf )}\Phi _2{\bf ({r},{R})=}\sum_{\beta
=n_{max}}^\infty g_\beta {\bf ({r})}f_\beta {\bf (R)}  \label{P}
\end{equation}
and
\begin{equation}
\hat F=\sum_\beta ^{n_{max}}|f_\beta \rangle \langle f_\beta |  \label{hatF}
\end{equation}

$n_{max}$ is a certain set of Coulomb quantum numbers chosen in such a way,
that $\Phi _2^d$ contains all the open channels and, eventually, a limited
number of closed ones. The sum from $n_{max}$ to infinity includes also the
integration over the continuous $\bar p$ momentum. The $\Phi _2^d$ component
describes the dynamics of Protonium formation and includes the corresponding
asymptotics of the three-body wavefunction :
\begin{eqnarray}
\lim_{r\rightarrow \infty }{\Phi _2^d({\bf {r},{R})=}}\sum_\beta f_\beta {%
{\bf {({R})\;}}}S_\beta {{\bf {\;}}}h_\beta ^{+}{{\bf {(r)}}}  \nonumber
\end{eqnarray}
Here $S_\beta $ are the S-matrix elements for the Protonium formation and $%
h_\beta ^{+}(r)$ are the outgoing electron waves in the channel with quantum
numbers $\beta$. The component $\Phi_2^p$ contains only closed channels.

At big $R$ the component $\Phi _2^d$ vanishes due to the Coulomb bound state
wave functions $f_\beta $. On the contrary, the contribution from $\Phi _2^p$ is
essential as far as it contains non vanishing terms coming from the p\=p
states in the continuum corresponding to the virtual excitations and
break-up. It is shown in Appendix that $\Phi _2^p$ actually describes the
effect of the long-range polarization which can be taken into account by
introducing in the elastic channel the local potential
\begin{equation}
V_{pol}(R)={\frac 12}{\frac{\alpha (R)}{R^4}}  \label{vpol}
\end{equation}
with $\alpha (R)$ ensuring that for $R>>r_B$ (H-Bohr radius) $\alpha
(R)\rightarrow -\alpha _d$ the H dipole polarizability. The following
approximation \cite{MM} for $\alpha (R)$ was found to be suitable in
practical calculations:
\[\alpha (R)={\frac 23}\left( R^5+\alpha _dR^4+2\alpha _dR^3+{\frac 32}\alpha
_d(R^2+R+{\frac 12})\right) e^{-2R}-\alpha _d \]

It is qualitatively shown in Appendix and proved by numerical calculations
that at distances $R\approx r_B$, $\Phi _2^d$ dominates and the main
contribution to the wavefunction comes from the channels with n=26-40, thus
we can choose $V_{pol}(R\ll r_B)\rightarrow 0.$

We will first consider the case of total $H\bar{p}$ angular momentum $L$
equal zero. The characteristic incident $\bar{p}$ energy, below which S-wave
dominates in the elastic channel will be determined later.

It was shown in \cite{AV,CV} that in the $L=0$ case, Protonium is primarily
produced in states with angular momentum $l=0$ and $l=1$. The physical
reason is that for the open channels with n=26-30, which dominate the
reaction amplitude, the electron is ejected with rather small momentum $k_e$%
, and the centrifugal barrier reduces the probability to find a slow
electron (and consequently Protonium) with high angular momentum.

The preceding results enable us to construct a model which, including a
limited number of channels, correctly describes the asymptotic behavior of
the three-body wavefunction. These channels dominate the reaction amplitude,
while closed channels corresponding to the continuous spectrum of Protonium
states, are taken into account by means of the polarization potential (\ref
{vpol}). In practical calculations we have included the Protonium channels
with principal quantum number n=10-40 and angular momentum $l=0,1$.
Numerical checks showed the stability of the results when  increasing
the number of included channels. The considered equation system has the
form:
\begin{eqnarray}
\left( -{\frac 1{2M}}\partial _R^2+{\frac{L(L+1)}{2MR^2}}%
+V_{cs}(R)+V_{pol}(R)-E+\varepsilon _B\right) \chi (R) &+&  \nonumber \\
+\quad \sum_{n,l}\int \phi _{1s}(r)W_{0l}(r,R)f_{n,l}(R)\hat \pi
g_{n,l}(r)dr &=&0  \label{FIRST}
\end{eqnarray}
\begin{eqnarray}
\sum_{n^{\prime }l^{\prime }}\left[ \left( -\frac 1{2m}\partial _r^2-\frac 1r%
+\frac{l(l+1)}{2mr^2}+E_n\right) \delta _{nn^{\prime },ll^{\prime }}+\hat \pi
U_{nn^{\prime },ll^{\prime }}(r)\hat \pi \right] g_{n^{\prime },l^{\prime
}}(r) &+&  \nonumber \\
+\quad \hat \pi \phi _{1s}(r)\int {f_{n,l}(R)W_{l0}(R,r)\chi (R)\;dR} &=&0
\label{SECOND}
\end{eqnarray}
Here :
\begin{eqnarray*}
V_{cs}(R)   &=& -\left( 1+{\frac 1R}\right) e^{-2R}            \cr
W_{0l}(r,R) &=&\frac{r_{<}^l}{r_{>}^{l+1}}\frac 1{\sqrt{2l+1}}\cr
U_{nn^{\prime },ll^{\prime }} &=&\int f_{n,l}{({\bf {R})}}f_{n^{\prime
},l^{\prime }}{{\bf ({R})\frac 1{|{R}-{r}|}d^3R}}  \cr
\hat \pi &=&1-\hat P\delta _{0l}
\end{eqnarray*}
and $E_n=-E+{\frac M{2n^2}}$


\subsection{Effective potential method}

The numerical solution of (\ref{FIRST}-\ref{SECOND}) is a rather difficult
task, as far as each of the coupled integro-differential equations includes
fast oscillating functions $\chi$ and $f_n$.

A way to overcome this difficulty is by means of the effective potential
method. This approach turned to be an efficient tool for both a qualitative
understanding and precise numerical treatment of the problem. The $H\bar p$
scattering observables are given by the first equation for antiproton
wavefunction $\chi $ (\ref{FIRST}). The effects of the remaining coupled
equations (\ref{SECOND}) are taken into account by transforming the system (%
\ref{FIRST}-\ref{SECOND}) into one equation for $\chi $ which contains a
complex nonlocal effective potential:
\begin{eqnarray}
\left( -{\frac 1{2M}}\partial _R^2+{\frac{L(L+1)}{2MR^2}}+V_{cs}(R)+\hat V%
_{eff}+V_{pol}(R)-E+\varepsilon _B\right) \chi (R)=0  \label{eqeff}
\end{eqnarray}
with
\begin{equation}
\hat V_{eff}=\sum_{nn^{\prime },ll^{\prime }}|f_{n,l}><\phi _{1s}|W_{0l}(r,R)%
\hat \pi \hat K_{nn^{\prime }}^{ll^{\prime }}(r,r^{\prime })\hat \pi
W_{l^{\prime }0}(r^{\prime },R^{\prime })|\phi _{1s}><f_{n^{\prime
},l^{\prime }}|  \label{Veff}
\end{equation}

Here $\hat K_{nn^{\prime }}^{ll^{\prime }}(r,r^{\prime })$ is the
Green-matrix of coupled equation system for electron wave-functions:
\begin{equation}  \label{K}
\hat{K}_{nn^{\prime }}^{ll^{\prime }}(r,r^{\prime }) = \left\{\left( -{\frac{%
1}{2m}}\partial^2_r +{\frac{l (l +1)}{2mr^2}}+ E_n-{\frac{1}{r}}%
\right)\delta _{nn^{\prime }}^{ll^{\prime }} +\hat\pi U_{nn^{\prime}}^{ll^{%
\prime }} \hat\pi\right\}^{-1}
\end{equation}
The whole problem is then splitted into two parts: to calculate the
effective potential and to solve the one-channel problem for antiproton
scattering in a complex nonlocal potential.

The benefits of such an approach are several. On one hand the Green function
(\ref{K}) is calculated by solving a coupled equation system for smooth
electron wavefunctions, while the fast oscillating $p\bar p$ wavefunctions
are explicitly introduced by the well-known Coulomb states. On the other hand
the effective potential~(\ref{Veff}) practically does not depend on the $%
\bar p$ incident energy in the domain $E_{\bar p} \ll 0.01$ a.u.
 ~\cite{AV,CV}.
The minimum energy of the ejected electron (from Pn state with n=30)
is about $0.02$ a.u. and for $\bar{p}$ energies less
than this value, the Green-matrix (\ref{K}) 
is not sensitive to incident antiproton energy. This means that,
once calculated for $E_{\bar p}=0$, 
the effective potential can be used in the whole energy range of 
interest and this radically simplifies the calculations.

\begin{figure}[tbp]
\vspace{+4.8cm} \epsfxsize=12cm \epsfysize=12cm
\par
\begin{center}
\mbox{\epsffile{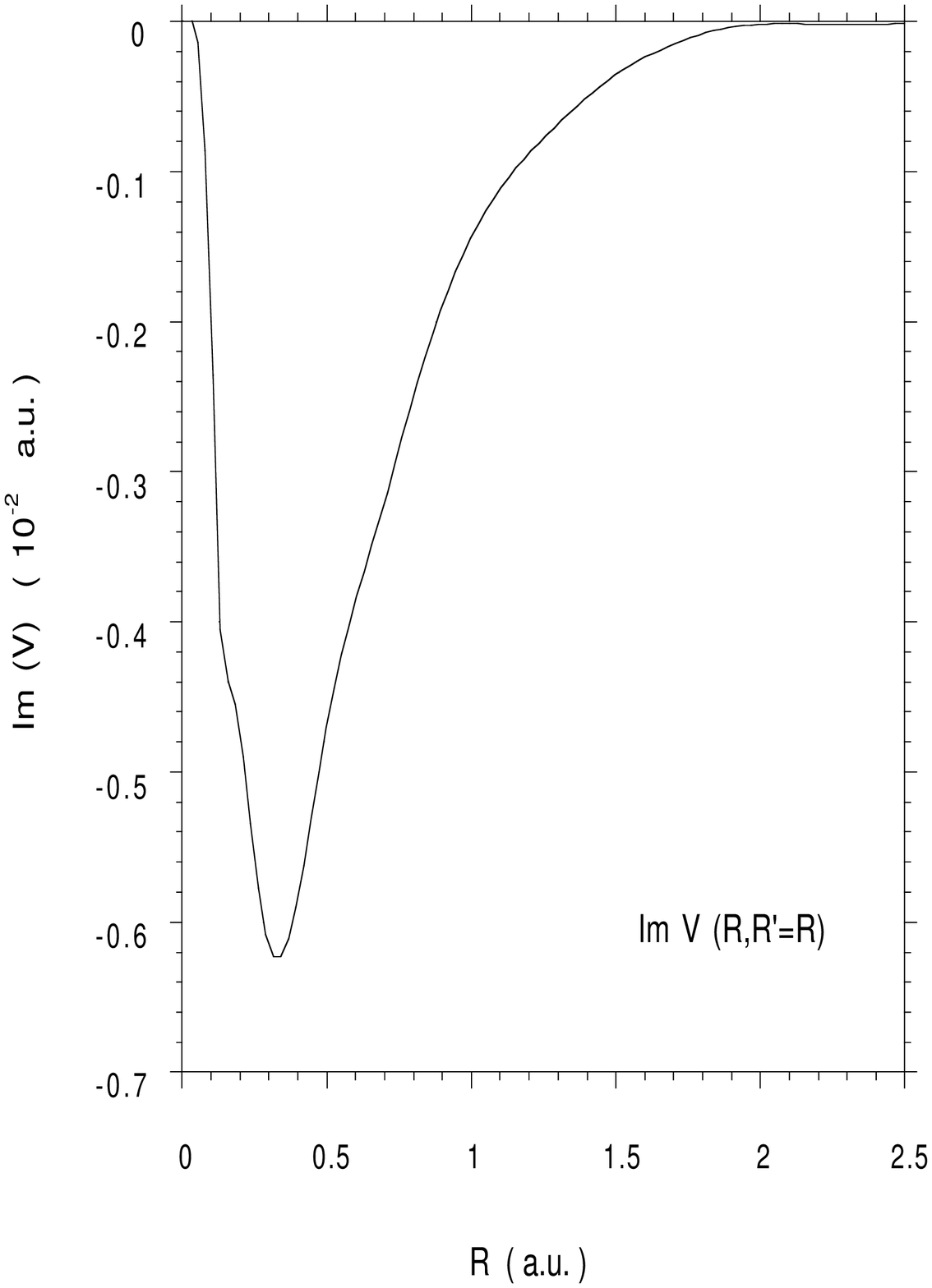}}
\end{center}
\caption{Imaginary part of effective potential $V_{eff}(R,R^{\prime}=R)$}
\label{Fig1}
\end{figure}

\begin{figure}[tbp]
\vspace{+4.8cm} \epsfxsize=12cm \epsfysize=12cm
\par
\begin{center}
\mbox{\epsffile{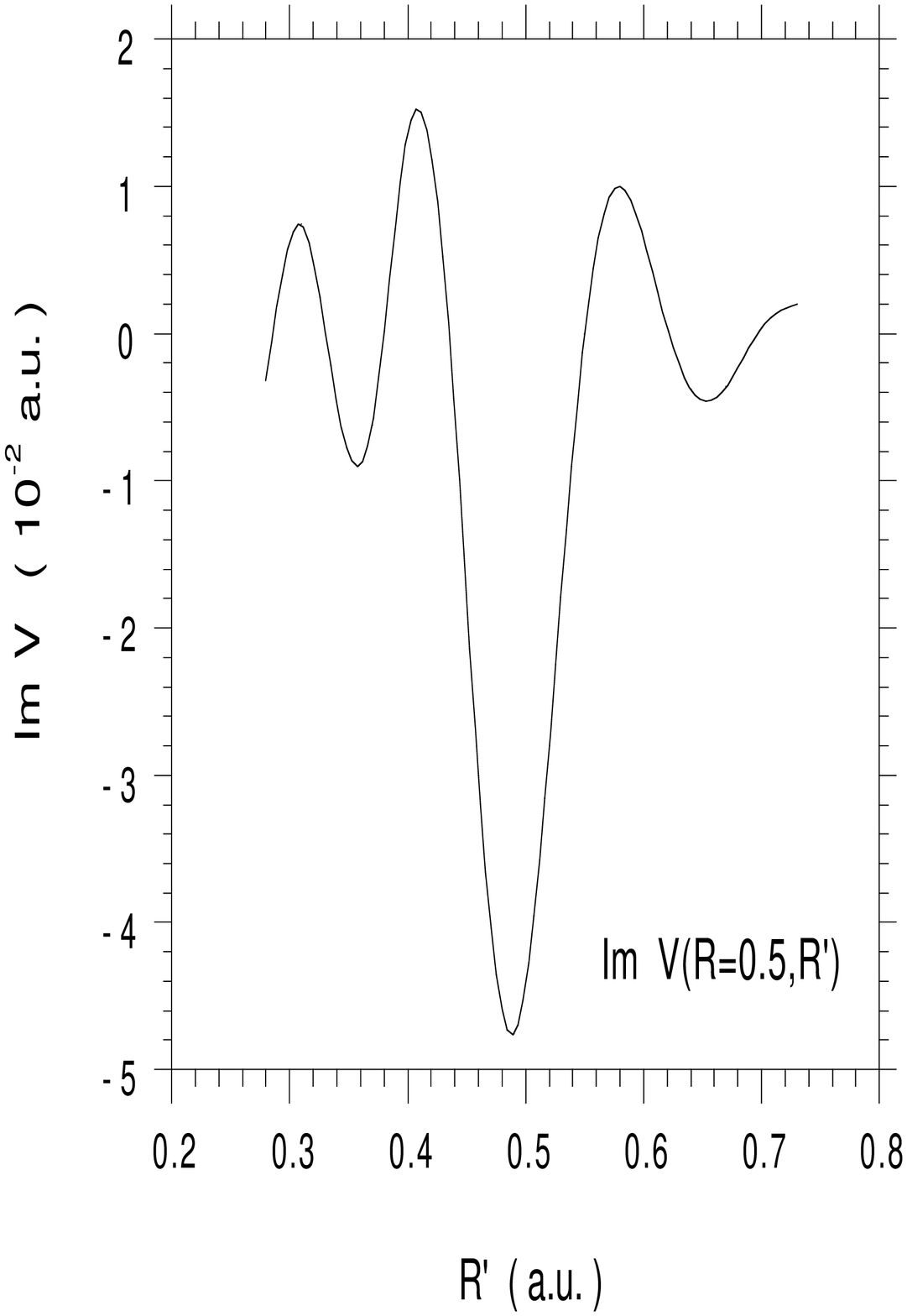}}
\end{center}
\caption{Imaginary part of effective potential $V_{eff}(R=0.5,R^{\prime})$}
\label{Fig2}
\end{figure}

\begin{figure}[tbp]
\vspace{+4.8cm} \epsfxsize=12cm \epsfysize=12cm
\par
\begin{center}
\mbox{\epsffile{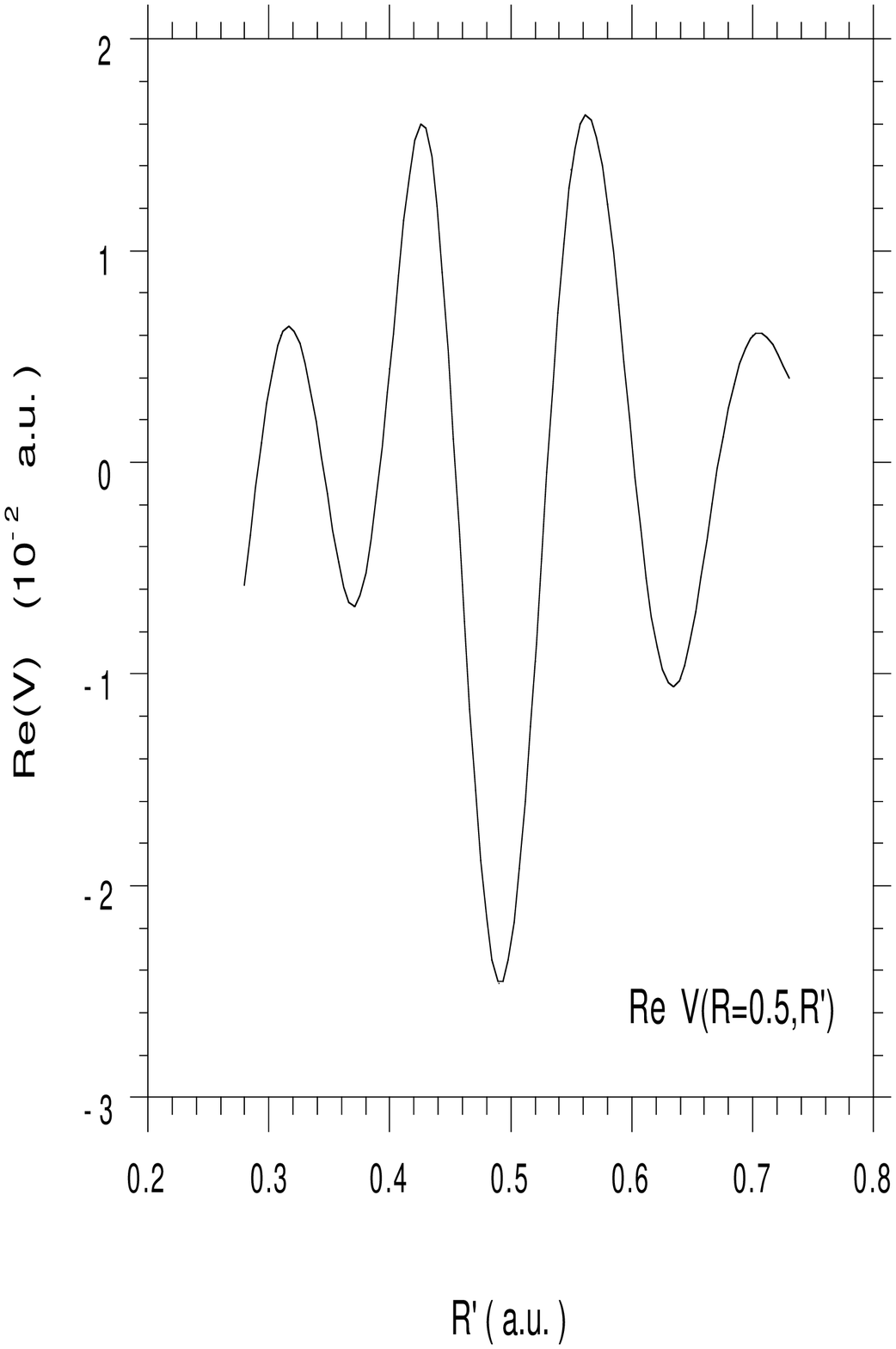}}
\end{center}
\caption{Real part of effective potential $V_{eff}(R=0.5,R^{\prime})$}
\label{Fig3}
\end{figure}

From a physical point of view it seems also more natural to analyze the
properties of the H\=p system in terms of a modified one channel problem.
The main features of the effective potential as they appear from our
calculations are the following:

\begin{enumerate}
\item  The imaginary part of $V_{eff}$ vanishes at distance $R\approx 1.8r_B$%
, which corresponds to the mean radius of the last Protonium open channel
(n=30). In Fig. \ref{Fig1} $Im[V_{eff}](R,R^{\prime })$ for $R^{\prime }=R$
is plotted as a function of R.

\item  The imaginary part of $V_{eff}(R,R^{\prime })$ is sharply peaked
around its diagonal $R^{\prime }=R$. Nevertheless for $R<r_B$, the
nonlocality range is of the same order as the antiproton wavefunction $%
\chi ({\bf {R})}$ oscillation period. The profile of $Im[V_{eff}](R,R^{%
\prime })$ for $R=0.5$ is shown on Fig. \ref{Fig2}.

\item  The profile of the real part of $V_{eff}$ is plotted in Fig. \ref
{Fig3}. Its nonlocality range is bigger than for the imaginary part. It
vanishes at $R\approx 3r_B$ and dominates over the polarization $V_{pol}$
and the Coulomb screened $V_{cs}$ potentials in the range $1<R<3r_B$
\end{enumerate}


\section{Results}

In this section we will present the main results obtained in the
coupled-channels model  and discuss the physical reasons of certain 
scattering observables behavior.

\subsection{Scattering observables}

The $H\bar{p}$ complex scattering length is found to be:
\[a= (-7.8 -i11.5) r_B \]
The corresponding elastic cross-section at zero energy is:
\[\sigma_{el}= 2426.4 r_B^2 \]
We remark the relatively big value, in the atomic scale, of the scattering
length imaginary part. Such value is a consequence of the long range
polarization forces. By switching off $V_{pol}$ in (\ref{eqeff}) the value
obtained is substantially reduced to $Im(a)= 0 .2 r_B$. The capital role of
the polarization forces in the low energy H\={p} dynamics will be discussed
in the next subsection.

\begin{figure}[tbp]
\vspace{+4.8cm} \epsfxsize=12cm \epsfysize=12cm
\par
\begin{center}
\mbox{\epsffile{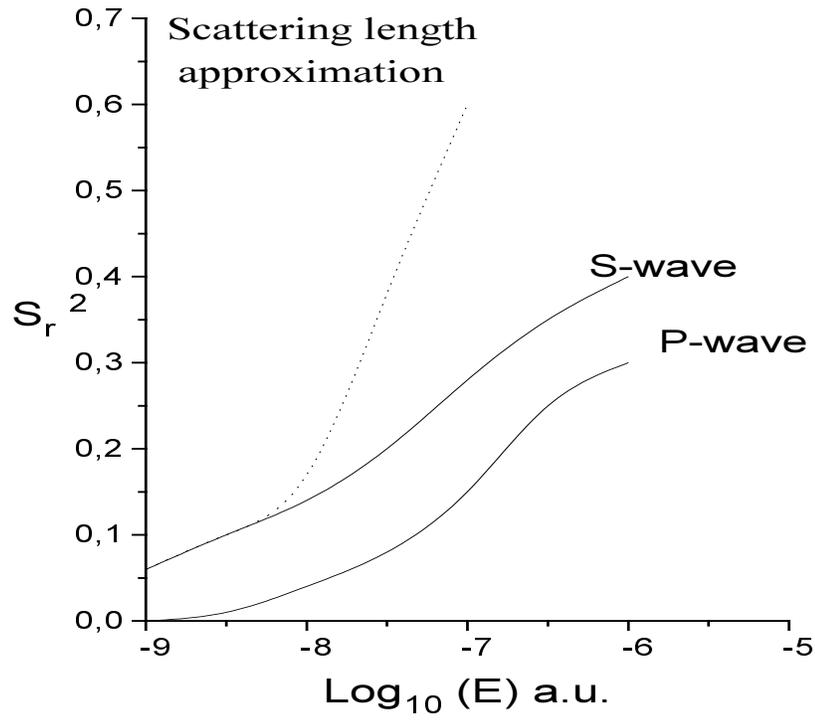}}
\end{center}
\caption{Inelasticity $S_r=1-|S|^2$ for the $H + \bar{p} \rightarrow Pn^{*}
+ e$ reaction}
\label{Fig4}
\end{figure}

We have calculated the energy dependence of the inelasticity $S^{2}_{r}$ for
several partial waves. The results are shown on Fig. \ref{Fig4} for $\bar{p}$
incident energies in the range from 0 to $10^{-6}$ a.u.. The inelasticity
turns to be less than 0.1 for $E_{\bar{p}}<10^{-8}$ a.u. and does not become
greater than $0.5$ in the energy domain of interest. One can also see in this
figure that the scattering length approximation is valid for energies less
than $10^{-8} a.u.$ The results for $l\neq 0$ have been calculated under the
assumption that the effective potential (\ref{Veff}) weakly depends on total
angular momentum L in the energy range of interest. As one can see, S-wave
dominates for $E_{\bar{p}} < 10^{-8} a.u.$.

\begin{figure}[tbp]
\vspace{+4.8cm}\epsfxsize=12cm\epsfysize=12cm
\par
\begin{center}
\mbox{\epsffile{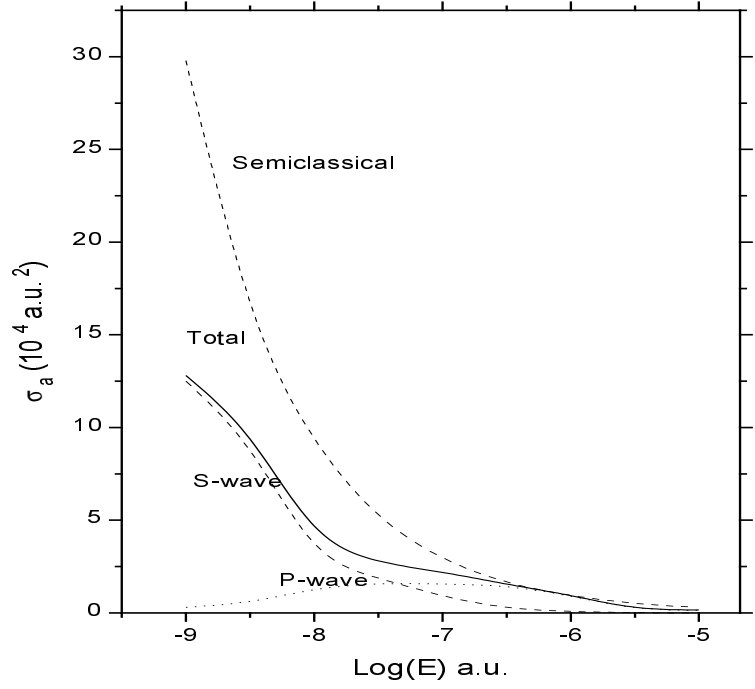}}
\end{center}
\caption{Annihilation cross section for $H + \bar{p} \rightarrow Pn^{*} + e$}
\label{Fig5}
\end{figure}

The total annihilation cross-section is shown on Fig. \ref{Fig5}. It follows
the $1/v$-law for $E_{\bar p}<10^{-8}$ a.u. and decreases nonmonotonously
for $E_{\bar p}>10^{-8}$ a.u. Such nonmonotonic behavior is originated by
the contribution of nonzero angular momentum partial waves, which is
explicitly seen on Fig. \ref{Fig4}.
It is interesting to compare this
cross-section with a semi-classical calculation \cite{Morgan1}
obtained under the following assumptions:
i) the $\bar{p}$ motion can be treated classically and ii) the annihilation
takes place with unit probability as soon as the $\bar{p}-H$
distance is smaller than a critical radius $R_c=0.64 r_B$.
The semi-classical cross-section,
shown on Fig. \ref{Fig5}, is approximately 2.5 times bigger
than our values for $E_{\bar p}<10^{-8}$ a.u. .
This indicates that the low-energy ${\bar p}H$
annihilation  is sensitive to the quantum dynamics of Protonium
formation and could hardly be reproduced with 
models in which the details of such dynamics are not taken into account.

\begin{figure}[tbp]
\vspace{+4.8cm}\epsfxsize=12cm\epsfysize=12cm
\par
\begin{center}
\mbox{\epsffile{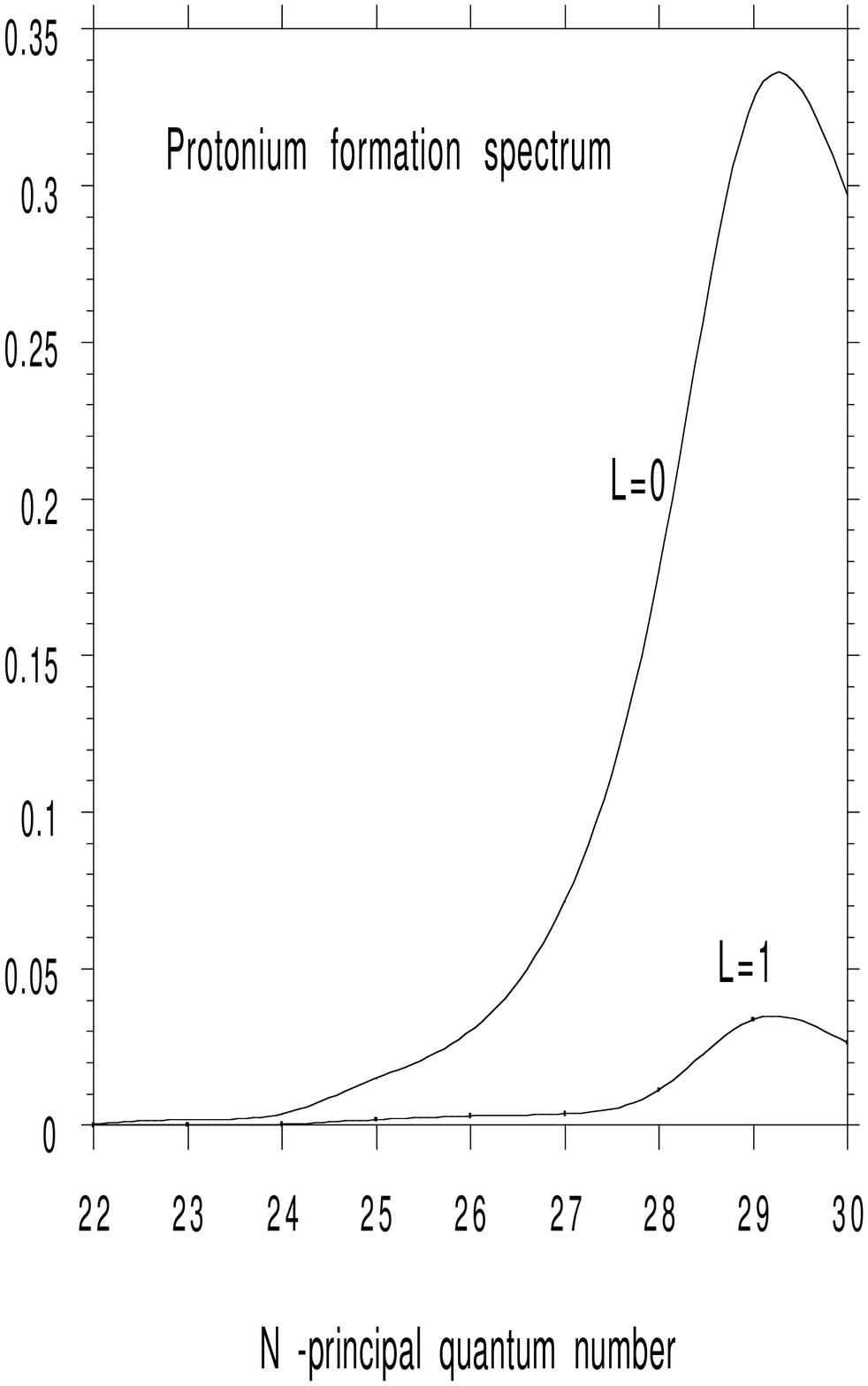}}
\end{center}
\caption{Protonium formation probabilities in states with different quantum
numbers}
\label{Fig6}
\end{figure}

The population of different Protonium states, calculated for energies $E_{%
\bar p}<10^{-8}a.u.$ is shown on Fig. \ref{Fig6}. Protonium is produced
primary in the S-states with principal quantum number $26<n<30$. The
P-states population does not exceed 15\% of the whole captured fraction of
antiprotons. These results confirm our qualitative estimation concerning the
channels which give the main contribution into the reaction amplitude in the
low-energy limit. It is worth to mention that Protonium S-states population
dominates only for $\bar p$ energies less than $10^{-8}$ a.u., while the
population of states with higher $l$ should increase with increasing energy
\cite{Protonium}.

We conclude this paragraph by emphasizing that the $H\bar p$ scattering
observables significantly change their behavior at $E_{\bar p}\sim 10^{-8}$
a.u., a characteristic energy for the reaction. We will demonstrate that
this behavior is determined by the presence of nearthreshold $H\bar p$ bound
and virtual states generated by the polarization potential.


\subsection{Nearthreshold metastable states}

The polarization potential is known to significantly modify the low energy
cross-sections of atomic reactions. It plays an essential role in the $H\bar{%
p}$ scattering. This potential produces a rich spectrum of $H\bar{p}$
weakly-bound and virtual states \cite{C}, which results from the long range
character of the polarization forces and the heavy (in atomic scale)
antiproton mass. Such states, being nearthreshold S-matrix singularities,
determine the energy dependence of the $H\bar{p}$ scattering cross-section.
The main properties of such states and their relation with the observables
are discussed in this subsection.

We first remark that the polarization potential $V_{pol}$ alone generates
several $\bar{p}$ weakly-bound states. The energy levels and mean radii of
several nearest to the threshold S-states produced by $V_{pol}$ alone
 are shown in Table \ref{Table1} (values marked by subscript II). 
These states are extremely prolonged and have very small binding
energies. By switching on the short range part of the interaction, i.e. the
complex nonlocal effective potential $V_{eff}$ and the screened Coulomb $%
V_{cs}$, the spectrum is modified and inelastic widths appear. Nevertheless,
the main features, small binding energy ($10^{-8}<E_{bound}< 10^{-3}$ a.u.) 
and big radius ($4<\bar{x}<27$) $r_B$, remain. 

\begin{figure}[tbp]
\vspace{+4.8cm} \epsfxsize=12cm\epsfysize=12cm
\par
\begin{center}
\mbox{\epsffile{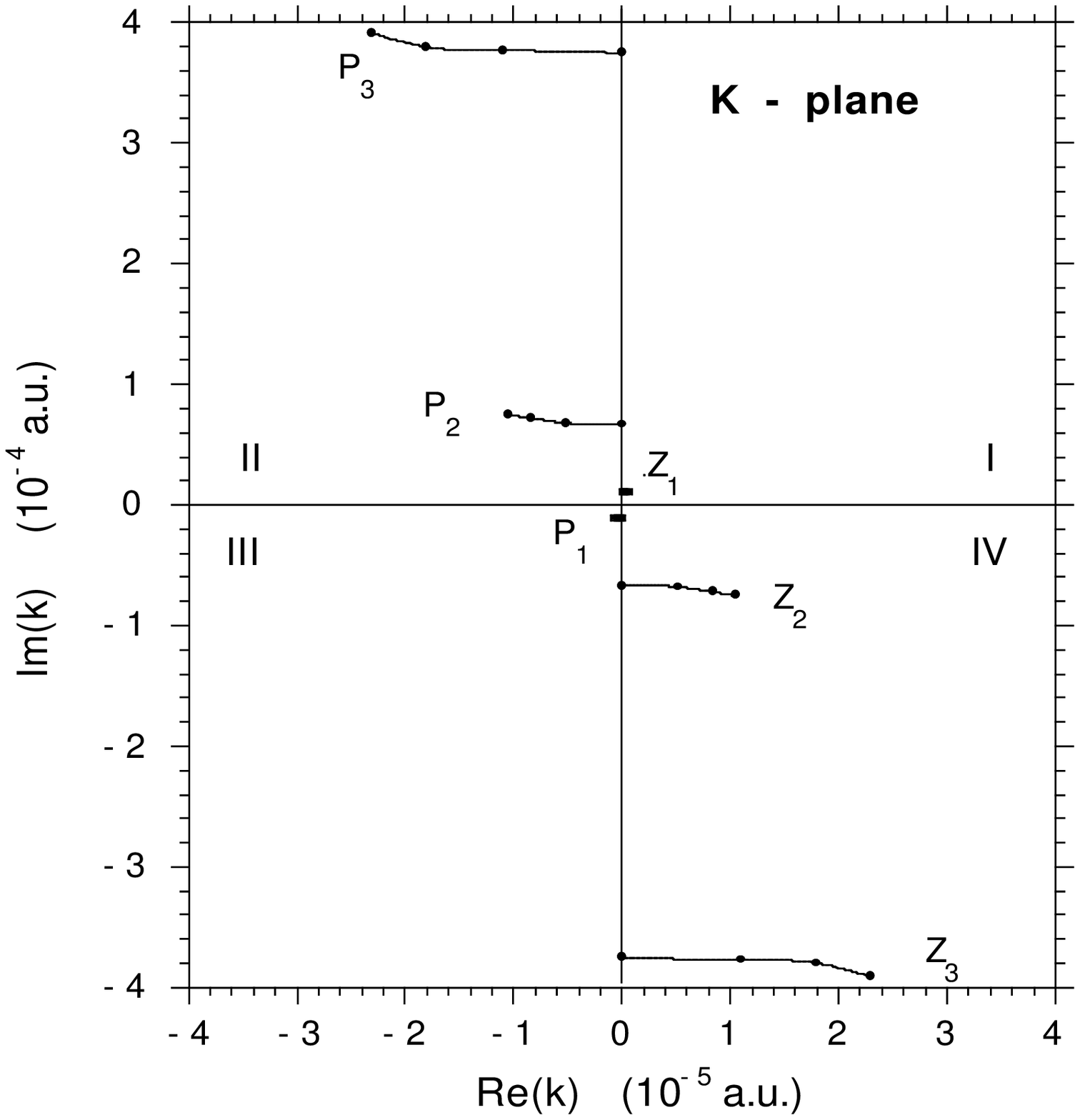}}
\end{center}
\caption{S-matrix nearthreshold zeros ($Z_i$) and poles ($P_i$) }
\label{Fig7}
\end{figure}

In the threshold vicinity the elastic S-matrix for L=0 is
dominated by its singularities and can be written in the form:
\begin{equation}  \label{Smatr}
|S|(k) = \prod_{i} \frac{|k+z_i|}{|k-z_i|}
\end{equation}
where $z_i$ are the S-matrix poles with $Re(z_i)<0$ due to $Im(V_{eff})<0$.
In Fig. \ref{Fig7} are shown the trajectories of several S-matrix poles ($P_i$)
and corresponding zeros ($Z_i$) as a function of the strength 
of the $V_{eff}$ imaginary part.
As it can be seen, the presence of the negative imaginary part in the effective
potential results in shifting the S-matrix zeros to the right into the IV
and I quadrants, with the corresponding symmetrical shift of S-matrix poles
into the II and III ones.
The position of the nearest to the origin S-matrix
zero (and pole) corresponds to an energy of $E_c \sim 10^{-8}$ a.u. and
plays a role of characteristic energy for the reaction ~(\ref{react}).
We notice however that this nearest to the threshold S-matrix singularity 
lies on the non-physical sheet, i.e. Re(k)$<0$ and Im(k)$<0$, 
and corresponds to a virtual state. Its
wave function has an exponentially incresing asymptotic 
and does not represent a physical state. 

As far as the usual definition of effective range can not be applied to the
$1/R^4$ polarization potential \cite{MM}, we introduce the characteristic
range of $H\bar{p}$ interaction as: $R_A=1/|k_0|$, where $k_0$ corresponds
to the position of the nearest to the threshold S-matrix singularity. One
can see, that for $k \ge k_0$ the scattering length approximation is no
longer valid and higher order terms in the scattering amplitude expansion
should be taken into account. With the result on Table I one gets
$R_A\sim 103 $ a.u. .

It is seen from ~(\ref{Smatr}) that for antiproton incident energies $E_{%
\bar{p}}\ll 10^{-8}$ a.u., $|k+z_i|\approx |k-z_i|$ and so $|S|\rightarrow 1
$. This explains why the inelasticity $S^{2}_{r}$ for $E_{\bar{p}}\ll
10^{-8} $ a.u. turns to be much less than unit. For $E_{\bar{p}}\ge10^{-8}$
a.u., and because there are several S-matrix zeros situated to the right
from $-z_1$, one has $|k+z_i|<|k-z_i|$ and the reaction probability
increases.

\begin{figure}[tbp]
\vspace{+4.8cm}\epsfxsize=12cm\epsfysize=12cm
\par
\begin{center}
\mbox{\epsffile{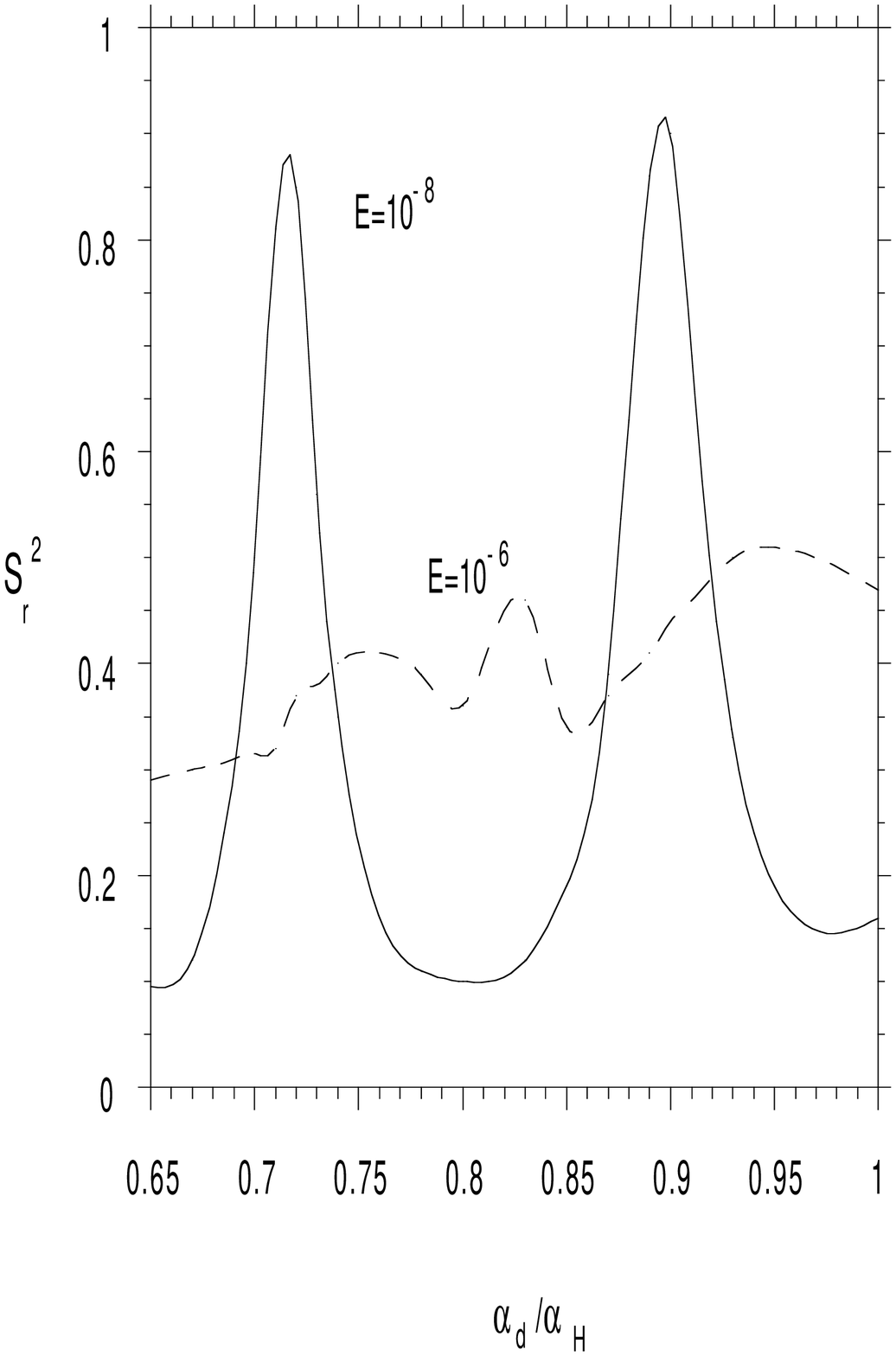}}
\end{center}
\caption{Inelasticity for reaction $H+\bar{p}\rightarrow Pn^{*}+e$ as a
function of the dipole polarizability $\alpha_d$}
\label{Fig8}
\end{figure}

To illustrate how the position of the nearthreshold S-matrix singularities
determines the low energy scattering, we have calculated the inelasticity as
a function of the dipole polarizability $\alpha_d$ for a fixed energy ($E_{%
\bar{p}}=10^{-8}$ and $E=10^{-6}$ a.u.). This function is plotted on Fig.
\ref{Fig8}. The strong oscillations between its maximum and minimum values
with decreasing $\alpha_d$ correspond to the motion of an S-matrix pole from
the II to the III k-plane quadrant, while the symmetric S-matrix zero moves
from the IV to the I quadrant. This means, that a weakly bound state becomes
virtual. As it is seen from ~(\ref{Smatr}), the inelasticity reaches its
maximum value when an S-matrix zero crosses the real k-axis.

This last result shows that sufficiently high accuracy of calculations is
required to obtain the scattering length value. In the same time the
reaction amplitude for energies $E_{\bar{p}}\gg 10^{-8}$ a.u. is less
sensitive to the exact position of the nearthreshold singularities and can
be more easily calculated. We estimate our accuracy in the scattering length
results to be about 30\%. This uncertainty appears mainly from approximation
used for $V_{pol}$ at short distances. To get more precise results one
should increase the number of accounted closed channels and take
into account the difference between ${\bf {\rho }}$ and ${\bf {R}}$ in (\ref
{Jfr}). Such corrections seem not to be important for understanding the
physics of the treated process and are beyond the scope of present paper.

We would like to emphasize that the nearthreshold character of mentioned
S-matrix poles and zeros is determined by the long-range polarization
potential. In the same time their exact position in complex k-plane can not
be obtained without a proper treatment of the Protonium formation dynamics.
In particular, the semiclassical black sphere condition  does not hold in the
energy domain $E_{\bar p}\le 10^{-6}$ a.u. . In terms of S-matrix analytical
properties the coupling with Protonium production channels produces
comparatively big (for the energy domain of interest) shifts of the real
part of the S-matrix zeros and reduces the inelasticity.


\subsection{Local approximation of the effective potential}

It was shown that the energy dependence of the reaction probability is
determined by the existence of several nearthreshold states generated mainly
by the long range polarization forces. This suggests the possibility to
obtain a local complex potential
which would be equivalent to the full $H\bar p$ interaction in the energy
range of interest. By equivalent we mean not only to reproduce the same
reaction probabilities but to support the same nearthreshold spectral
structure as well.

We search for such equivalent local complex potential as a sum of three
different terms
\[
V_{loc} (R) = V_s(R) +V_{cs}(R) +V_{pol}(R)
\]
$V_{cs}(R)$ and $V_{pol}(R)$ being respectively the Coulomb screened and
polarization potential used in the previous section and $V_s$ a local short
rang part to be determined. It was assumed to have the form:
\begin{eqnarray}
V_{s}(R) &=&\left\{
\begin{array}{cll}
-V_1\;e^{-\alpha_1\left({\frac{R}{r_B}}\right)} -i \; W_1 \;
e^{-\beta_1\left({\frac{R}{r_B}}\right)} & \mbox{if} & R < R_c \\
-i\;W_2\; e^{-\beta_2\left({\frac{R}{r_B}}\right) } & \mbox{if} & R\geq R_c
\end{array}
\right.
\end{eqnarray}
and a satisfactory fit is obtained with the following parameter values: $%
V_1=0.572$, $W_1=W_2=0.040$, $\alpha_1=1.20$, $\beta_1= \beta_2=3.20$ and $%
R_c=2r_B$.

In Table \ref{Table2} the results of calculations in nonlocal effective
potential and mentioned above local approximation are compared. They agree
within few percent accuracy in the energy range $0.5\;10^{-9}-0.5\;10^{-6}$
a.u.


\subsection{Hydrogen-Antihydrogen interaction}

The results obtained for $H\bar{p}$ interaction can be used for a
qualitative treatment of different atom-antiproton (A\={p}) and
atom-antiatom (A\={A}) system.

It is of particular interest to estimate the $H\bar H$ annihilation
cross-section, and thus to examine the reaction:
\begin{equation}
H+\bar H\rightarrow Pn^{*}+(e^{+}e^{-})  \label{Hbar}
\end{equation}
The $H\bar H$ system interacts at long distances via a dipole-dipole
potential $V_{dd}\sim -6.5/R^6$. This potential also generates a spectrum of
nearthreshold states. Some of them with L=0 are shown in Table \ref{Table3}.
In analogy with the (\ref{react}) case one can expect that the corresponding
S-matrix singularities will determine the (\ref{Hbar}) reaction dynamics.

\begin{figure}[tbp]
\vspace{+4.8cm} \epsfxsize=12cm\epsfysize=12cm
\par
\begin{center}
\mbox{\epsffile{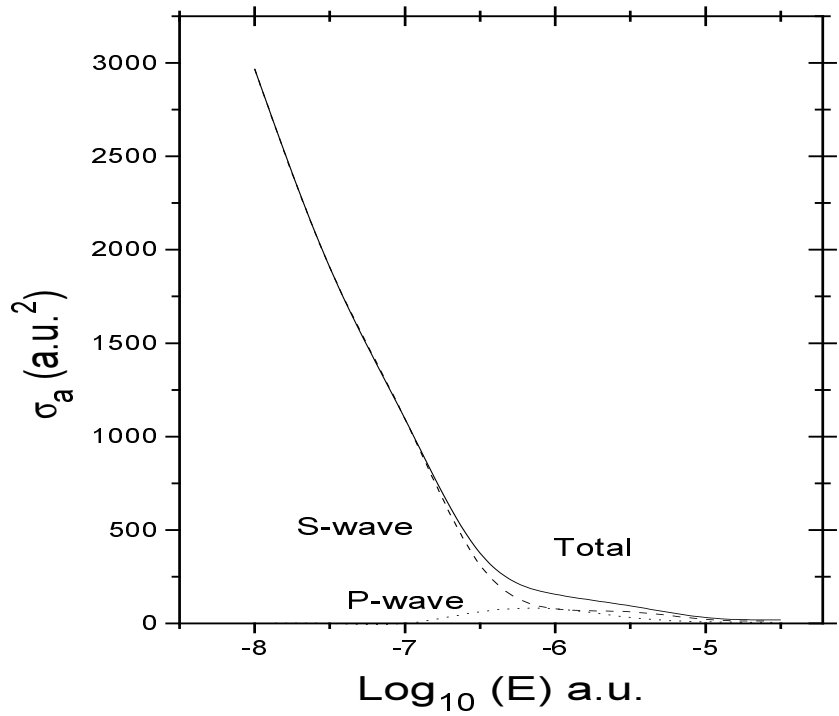}}
\end{center}
\caption{Probability of reaction $H + \bar{H} \rightarrow Pn^{*} + (e^+e^-)$}
\label{Fig9}
\end{figure}

A qualitative estimation of the $H\bar{H}$ potential can be obtained by
adding to the same short-range part as in $H\bar{p}$ case the dipole-dipole
long-range tail $V_{dd}$. The reaction ~(\ref{Hbar}) cross-section
calculated in such a way is shown on Fig. \ref{Fig9}. The characteristic
energy for this reaction was found to be $\sim 10^{-5}$ a.u. , corresponding
to the position of the nearest to the threshold S-matrix singularity
(virtual state with energy $-7.8\;10^{-6}$ a.u.)

A similar treatment can be used to estimate the inelasticity energy
dependence for different A\={p} or A\={A} systems in the low energy limit.
For such a purpose one has to find the nearest to the threshold S-matrix
singularity, generated by polarization potential. The necessary condition
for the validity of such a qualitative approach is that the characteristic
range $R_{A}$ of the A\={p} or A\={A} long-range interaction should be much
greater than the inelastic range $r_{A}$.

As it is seen from (\ref{Veff}) the inelastic range is mainly determined by
the mean radius of the last Protonium state open channel and thus given by:
\begin{eqnarray}
{\frac{M_A}{2n^2}} &=&I_A  \nonumber \\
r_A &\approx &{\frac{2n^2}{M_A}}={\frac 1{I_A}}  \nonumber
\end{eqnarray}
where $M_A$ and $I_A$ are the A\=p reduced mass and the first ionization
potential respectively, n is the principal quantum number of the last open
channel. A similar estimation for the A\=A inelastic range $r_{A\bar A}$ can
be obtained, if we take into account that Positronium is produced in this
collision:
\begin{eqnarray}
{\frac{M_{A\bar A}}{2n^2}} &=&2I_A-\varepsilon _{Ps}  \nonumber  
\\
\varepsilon _{Ps} &=&I_H/2  \nonumber \\
r_A &\approx &{\frac{2n^2}{M_{A\bar A}}}={\frac 1{(2I_A-I_H/2)}}  \nonumber
\end{eqnarray}
Here $M_{A\bar A}$ is the reduced mass of the A\=A system, $\varepsilon
_{Ps} $ is the Positronium ground state energy.

Like in the $H\bar p$ case, the presence of nearthreshold
virtual states may considerably increase the characteristic range of A\=p or
A\=A interaction. However it can be interesting to have a simple approximation
of this range in the aim of comparison with preliminary estimations
(see also \cite{SWS}).
This is provided by the semiclassical condition
for the number N of states :
\[\int {\sqrt{2M_AV_{pol}^A(R)}dR}\approx \pi N  \]
This condition may be rewritten as follows:
\[{\frac{R_A}{r_B}}\approx \pi N \]
For the A\=p case we obtain:
\begin{eqnarray}
R_A\sim \left\{
\begin{array}{cll}
\sqrt{2M_AC_4^A} & \mbox{if} & L=0 \\
\sqrt{2M_AC_4^A/L(L+1)} & \mbox{if} & L>0
\end{array}
\right.
\end{eqnarray}
while for A\=A one has:
\begin{eqnarray}
R_A\sim \left\{
\begin{array}{cll}
\sqrt[4]{2M_{A\bar A}C_6^A} & \mbox{if} & L=0 \\
\sqrt[4]{2M_{A\bar A}C_6^A/L(L+1)} & \mbox{if} & L>0
\end{array}
\right.
\end{eqnarray}
$C_4^A$ and $C_6^A$ being the atom charge-dipole and dipole-dipole van der
Waals constants. Finally, we get the following ratio of inelastic and
polarization range:
\begin{eqnarray}
{\frac{r_A}{R_A}}=\left\{
\begin{array}{cll}
{\frac 1{I_A\sqrt{2M_AC_4^A}}} & \mbox{if} & L=0 \\
{\frac{\sqrt{L(L+1)}}{I_A\sqrt{2M_AC_4^A}}} & \mbox{if} & L>0
\end{array}
\right.  \label{Ap}
\end{eqnarray}
for atom-antiproton, and:
\begin{eqnarray}
{\frac{r_A}{R_A}}=\left\{
\begin{array}{cll}
{\frac 1{(2I_A-I_H/2)\sqrt[4]{2M_{A\bar A}C_6^A}}} & \mbox{if} & L=0 \\
{\frac{\sqrt[4]{L(L+1)}}{(2I_A-I_H/2)\sqrt[4]{2M_{A\bar A}C_6^A}}} & %
\mbox{if} & L>0
\end{array}
\right.  \label{AA}
\end{eqnarray}
for atom-antiatom interaction.

The ratios (\ref{Ap}-\ref{AA}), calculated for a wide range of different
atoms, turns to be much smaller than unit in case L=0. In particular, for
He, the less polarizable atom, they are $\sim 0.02$ for $He$$\bar{p}$, and $%
\sim 0.05$ for $He\bar{He}$. The polarization range dominates over the
inelastic one in the partial waves up to $L\sim10$ for $He\bar{p}$, and $%
L\sim4$ for $He\bar{He}$. These values of $L$ characterize the maximum
angular momentum, which makes possible the existence of extended
polarization states.

\section{Conclusion}

A coupled channels model describing the $H\bar p$ system at energies
less than $10^{-6}$ a.u. has been developed.
The results thus obtained substantially differ from the low energy
extrapolations of the black sphere model and other classical
or semiclassical approaches.
They show that such a low energy requires a quantum mechanical
treatment in which the dynamics of the Protonium formation is properly
taken into account.

The effective $H\bar{p}$ optical potential has been calculated in the
framework of the coupled channels model.
In this framework, the $H\bar p$ scattering length and zero energy elastic
cross section were
found to be $a=(-7.8-i11.5)r_B$ and $\sigma _{el}=2426.4r_B^2$ respectively.
The $H\bar p$ inelastic cross section has been calculated in the energy
range from $10^{-9}$ to $10^{-6}$ a.u.. It follows the $1/v$ behavior up to
energies $\sim 10^{-8}$ a.u. where the scattering length approximation is
valid. The inelasticity turned to be much smaller than the black sphere
model predictions.

The Protonium formation spectrum for the energies less than $10^{-8}$ a.u.
has been calculated.
We have shown that the population of S-states with principal
quantum number from 26 to 30 acounts for 75\% of the total captured fraction.

The reaction dynamics is found to be determined by the existence of several
nearthreshold states. Such states are produced by the long-range
polarization potential and are shifted in the complex momentum plane by the
coupling with Protonium formation channels. The $H\bar{p}$ scattering length
appears to be very sensitive to the position of the mentioned singularities
and requires accurate calculations.

A local approximation of the effective potential has been proposed for
further applications. It reproduces the scattering observables in the
considered energy range and has the same nearthreshold spectral properties.

A qualitative extension of this approach to more general systems (atom-$\bar{%
p}$ and atom-antiatom) has been discussed.

The results discussed in this work have been obtained
within an approximate model.
In view of them and motivated by the futur project
of storing antimater at CERN it would be interesting to check the
validity of the different approximations by developing more accurate
treatments including
an exact solution of the three body problem.


\section{Acknowledgments}

The authors would like to thank I.S. Shapiro for suggesting the problem. One
of the authors (A.V.) would like to thank D. Morgan for useful discussions.

\appendix

\section{Appendix}

The aim of this appendix is to find the dominant channels in the expansion (%
\ref{phi2exp}) of the three-body wavefunction.
We first analyze the behavior of the component $\Phi _2^p$ at distances $%
R\gg r_B$. The equations system for $\Phi_2^d$ and $\Phi_2^p$, in terms of
the projection operators $\hat P$ (\ref{hatP}) and $\hat F$ reads: (\ref
{hatF}):
\begin{eqnarray}
(\hat H_{p\bar p}+\hat P\hat W_{e\overline{p}}\hat P-E_{\bar p})|\Phi
_1\rangle +\hat P\hat W_{e\overline{p}}(1-\hat P)\hat F|\Phi _2^d\rangle &+&
\nonumber   \\
+\;\hat P\hat W_{e\overline{p}}(1-\hat P)(1-\hat F)|\Phi _2^p\rangle &=&0
\label{C1}
\end{eqnarray}
\begin{eqnarray}
\left( \hat H_{ep}+\hat H_{p\bar p}+\hat F(1-\hat P)\hat W_{e\overline{p}}(1-%
\hat P)\hat F-E\right) |\Phi _2^d\rangle &+&  \nonumber  \\
+\hat F(1-\hat P)\hat W_{e\overline{p}}(1-\hat P)(1-\hat F)|\Phi _2^p+(1-%
\hat P)\hat F\hat W_{e\overline{p}}\hat P|\Phi _1\rangle &=&0  \label{C2d}
\end{eqnarray}
\begin{eqnarray}
\left( \hat H_{ep}+\hat H_{p\bar p}+(1-\hat F)(1-\hat P)\hat W_{e\overline{p}%
}(1-\hat P)(1-\hat F)-E\right) |\Phi _2^p\rangle &+&  \nonumber
\\
+(1-\hat F)(1-\hat P)\hat W_{e\overline{p}}\hat P\hat F|\Phi _2^d+(1-\hat P%
)(1-\hat F)\hat W_{e\overline{p}}\hat P|\Phi _1\rangle &=&0  \label{C2p}
\end{eqnarray}

By taking into account that at big $R$ the projection operator $\hat F$
vanishes, equations (\ref{C1}-\ref{C2p}) simplify into the following system,
valid for $R\gg r_B$:
\begin{eqnarray}
\left( \hat H_{p\bar p}+\hat P\hat W_{e\overline{p}}\hat P-E_{\bar p}\right)
|\Phi _1\rangle +\hat P\hat W_{e\overline{p}}(1-\hat P)|\Phi _2^p\rangle &=&0
\label{A2} \\
\left( \hat H_{p\bar p}+\hat H_{ep}+(1-\hat P)\hat W_{e\overline{p}}(1-\hat P%
)-E\right) |\Phi _2^p\rangle +(1-\hat P)\hat W_{e\overline{p}}\hat P|\Phi
_1\rangle &=&0  \label{A3}
\end{eqnarray}

This system can be solved with respect to $\Phi_2^p$ and gives for the
component $\Phi_1$:
\begin{eqnarray}
\left( \hat H_{p\bar p}+\hat P\hat W_{e\overline{p}}\hat P-E_{\bar p}\right)
|\Phi _1\rangle +\hat P\hat W_{e\overline{p}}(1-\hat P)\hat G_{pol}(1-\hat P)%
\hat W_{e\overline{p}}\hat P|\Phi _1\rangle =0  \label{E1}
\end{eqnarray}
in which
\begin{eqnarray}
\hat G_{pol}=-\left( \hat H_{p\bar p}+\hat H_{ep}+(1-\hat P)\hat W_{e%
\overline{p}}(1-\hat P)-E\right) ^{-1}  \label{Gpol}
\end{eqnarray}

The last term in equation (\ref{E1}) is the polarization long-range
interaction:
\begin{equation}
\hat V_{pol}=\langle \phi _{1s}|\hat W_{e\overline{p}}(1-\hat P)\hat G%
_{pol}(1-\hat P)\hat W_{e\overline{p}}|\phi _{1s}\rangle  \label{Vp}
\end{equation}
The asymptotics of the Green-function $\hat G_{pol}$ at $R,R'\gg r_B$ is:
\[G_{pol}(R,R^{\prime },r,r^{\prime })=
\sum_{\alpha}{1\over 2p_\alpha}
\left( e^{-p_\alpha |R-R^{\prime }|}-S_\alpha e^{-p_\alpha
(R+R^{\prime })}\right) \phi _\alpha (r)\phi _\alpha (r^{\prime }) \]
where $p_\alpha =\sqrt{2M(|E-\varepsilon _\alpha |)}$ and $\alpha $ is a set
of spherical Coulomb quantum numbers. If we take into account that at big R,
$\hat V_{pol}$ acts on the very slowly changing function $\chi (R)$
(the oscillation period of $\chi$  for $R\gg r_B$
is indeed much greater than $r_B$)
we can substitute $G_{pol}$
in (\ref{Vp}) by the following expression:
\[G_{pol}(R,R^{\prime},r,r^{\prime })=\delta \left( R-R^{\prime
}\right) \sum_{\alpha }{\frac{\phi _\alpha (r)\phi _\alpha
(r^{\prime })}{E-\varepsilon _\alpha}} \]

By keeping terms up to $1/R^4$ we obtain the well-known charge-dipole
potential asymptotic behavior:
\[V_{pol}(R\gg r_B)=-{\frac{\alpha _d}{2R^4}} \]
\begin{equation}
\alpha _d=-2\sum_{\alpha \ne (1s)}\langle \phi _{1s}|\hat d|\phi {\alpha }%
\rangle {\frac 1{\varepsilon _B-\varepsilon _\alpha }}\langle \phi _\alpha |%
\hat d|\phi _{1s}\rangle  \label{alphad}
\end{equation}
Here $\hat d$ stands for the dipole momentum operator. One recognizes in (%
\ref{alphad}) the expression for the Hydrogen dipole polarizability, $\alpha
_d=9/2$. We can thus conclude that the contribution of $\Phi_2^p$ at big
distances $R$ can be taken into account by introducing in the elastic
channel the polarization charge-dipole potential.

To qualitatively estimate the contribution of different channels at
distances $R\approx r_B$ we first obtain the solution of (\ref{C1}) in the
distorted wave approximation. Component $\Phi _{1s}^0$ , obtained by
neglecting the coupling to other components, is $\Phi _1^0=\phi _{1s}\chi ^0$
with $\chi ^0$ satisfying the equation:
\[(-{\frac 1{2M}}\partial _R^2+V_{cs}(R)-E+\varepsilon _B)\chi ^0=0 \]
The contributions of components $\Phi _2^d$ and $\Phi _2^p$ are characterized
by the integrals:
\begin{eqnarray}
&&\ \langle \chi ^0\hat P|\hat W_{e\overline{p}}(1-\hat P)|\hat F\nonumber\\
&&  \label{Int} \\
&&\ \langle \chi ^0\hat P|\hat W_{e\overline{p}}(1-\hat P)|(1-\hat F)
\nonumber
\end{eqnarray}

An estimation of integrals (\ref{Int}) can be obtained if we take into
account the semiclassical character of the wavefunction $\chi ^0$ and the
Coulomb wavefunctions at $R\approx r_B$ in expansions (\ref{D}) and (\ref{P}%
). We are dealing with an integral of fast-oscillating functions which has
significant values only if there exist stationary phase points inside the
integration region. The equation for such stationary phase points is:
\[
\left( {\frac 1R}+1\right) e^{-2R}={\frac 1R}-{\frac M{2n^2}}
\]
It can be shown that there are no stationary phase points for $\Phi _2^p$,
while the contribution of $\Phi _2^d$ at the distance $R\approx r_B$ is
mainly exhausted by Protonium states with principal quantum number $26<n <40$
(see \cite{AV,CV}).

We have, in conclusion, that in the energy domain of interest and large
internucleon distances the contribution of $\Phi _2^p$ is the only important
one and can be taken into account by introducing the polarization potential (%
\ref{Vp}), while at the distances $R\approx r_B$ the component $\Phi _2^d$
dominates, and can be described by a limited number of channels. These
qualitative arguments are important for construction of the first
approximation, and should be proved by further numerical calculations.



\newpage
\begin{table}[tbp]
\[
\begin{array}{|c|c|c|}
\hline
E_I & E_{II} & \bar{x}_{II} \\ \hline
& -5.1\;10^{-8} +i\; 7\;10^{-9} &  \\
-4.2\;10^{-7} & -2.5\;10^{-6} -i\; .2\;10^{-7} & 27.0 \\
-3.6\;10^{-5} & -7.0\;10^{-5} -i\;8.4\;10^{-6} & 11.3 \\
-2.6\;10^{-4} & -4.1\;10^{-4} -i\;3.2\;10^{-5} & 7.3 \\
-9.2\;10^{-4} & -1.5\;10^{-3} -i\;8.6\;10^{-5} & 5.3 \\
-2.3\;10^{-3} & -4.2\;10^{-3} -i\;2.0\;10^{-4} & 4.2 \\ \hline
\end{array}
\]
\caption{Energies, Auger widths and mean radii (a.u) of L=0 $H\bar{p}$
states. We denote by index I the results in the $V_{pol}$ alone and by index
II those obtained with the full interaction ($V_{pol}+V_{cs}+V_{eff}$) }
\label{Table1}
\end{table}

\vspace{2.cm}
\begin{table}[tbp]
\[
\begin{array}{|c|c|c|c|c|c|}
\hline
E_{\bar{p}}\quad$(a.u.)$ & S_r^2\quad(I) & S_r^2\quad(II) & S\quad(I) & S
\quad(II) &  \\ \hline
0.5\;10^{-9} & 0.043 & 0.043 & 0.978+ i0.01 & 0.978+ i0.014 &  \\
0.5\;10^{-8} & 0.12 & 0.122 & 0.937+i0.013 & 0.936+i0.021 &  \\
0.5\;10^{-7} & 0.266 & 0.266 & 0.836-i0.185 & 0.836-i0.177 &  \\
0.5\;10^{-6} & 0.42 & 0.425 & 0.023-i0.756 & 0.034-i0.757 &  \\ \hline
\end{array}
\]
\caption{Inelasticity ($S_r^2$) and S-matrix (S) values calculated in the
full effective potential (index I) and in its local approximation
(index II) at different energies ($E_{\bar{p}}$) }
\label{Table2}
\end{table}

\vspace{2.cm}
\begin{table}[tbp]
\[
\begin{array}{|c|c|c|}
\hline
E_I & E_{II} & \bar{x}_{II} \\ \hline
-7.8\;10^{-6} & -6.1\;10^{-6} +i\; 1.8\;10^{-5} &  \\
-1.9\;10^{-4} & -4.3\;10^{-4} -i\;2.2\;10^{-4} & 4.6 \\
-2.9\;10^{-3} & -5.2\;10^{-3} -i\;1.2\;10^{-3} & 2.8 \\
-1.1\;10^{-2} & -2.9\;10^{-2} -i\;8.4\;10^{-3} & 1.5 \\
-3.3\;10^{-2} & -5.8\;10^{-3} -i\;9.2\;10^{-3} & 1.3 \\ \hline
\end{array}
\]
\caption{Energies, Auger widths and mean radii (a.u) of L=0 $H\bar{H}$
states. We denote by index I the results in the $V_{pol}$ alone and by index
II those obtained with the full interaction ($V_{pol}+V_{cs}+V_{eff}$) }
\label{Table3}
\end{table}


\newpage

\newpage

\newpage

\newpage

\newpage

\newpage

\newpage

\newpage

\newpage

\end{document}